\documentclass[aps,reprint,showpacs,floatfix,a4paper]{revtex4-1}

\usepackage{graphicx}
\usepackage{amsmath}
\usepackage{amssymb}
\usepackage{amsfonts}
\usepackage{color}
\usepackage{relsize}
\usepackage{mathrsfs}
\usepackage{txfonts}
\usepackage[T1]{fontenc}
\usepackage{bm}
\usepackage[final,kerning,spacing,babel]{microtype}
\usepackage[pdfpagemode=UseNone,pdfstartview=FitH]{hyperref}
\usepackage{epsfig}
\usepackage{tabularx}

\begin{document}

\title{Probing deformed commutators with macroscopic harmonic oscillators}

\author{Mateusz~Bawaj$^{1,2}$,
Ciro~Biancofiore$^{1,2}$,
Michele~Bonaldi$^{3,4}$,
Federica~Bonfigli$^{1,2}$,
Antonio~Borrielli$^{3,4}$,
Giovanni~Di Giuseppe$^{1,2}$,
Lorenzo~Marconi$^{5}$,
Francesco~Marino$^{6,7}$,
Riccardo~Natali$^{1,2}$,
Antonio~Pontin$^{5,6}$,
Giovanni~A.~Prodi$^{4,8}$,
Enrico~Serra$^{4,8,9,10}$,
David~Vitali$^{1,2}$,
Francesco~Marin$^{5,6,11,\star}$}

\affiliation{$^{1}$School of Science and Technology, Physics Division, University of Camerino, via Madonna delle Carceri, 9, I-62032 Camerino (MC), Italy,
$^{2}$INFN, Sezione di Perugia, Italy,
$^{3}$Institute of Materials for Electronics and Magnetism, Nanoscience-Trento-FBK Division, I-38123 Povo (TN), Italy,
$^{4}$Istituto Nazionale di Fisica Nucleare (INFN), Trento Institute for Fundamental Physics and Application, I-38123 Povo (TN), Italy,
$^{5}$Dipartimento di Fisica e Astronomia, Universit\`a di Firenze, Via Sansone 1, I-50019 Sesto Fiorentino (FI), Italy,
$^{6}$INFN, Sezione di Firenze, Via Sansone 1, I-50019 Sesto Fiorentino (FI), Italy,
$^{7}$CNR-Istituto Nazionale di Ottica, Largo E. Fermi 6, I-50125 Firenze, Italy,
$^{8}$Dipartimento di Fisica, Universit\`a di Trento, I-38123 Povo (TN), Italy,
$^{9}$Centre for Materials and Microsystem, Fondazione Bruno Kessler, I-38123 Povo (TN), Italy,
$^{10}$Dept. of Microelectronics and Computer/ECTM/DIMES Technology Centre, Delft University of Technology, Feldmanweg 17, 2628 CT Delft, PO Box 5053, 2600 GB Delft, The Netherlands,
$^{11}$European Laboratory for Non-Linear Spectroscopy (LENS), Via Carrara 1, I-50019 Sesto Fiorentino (FI), Italy.
$\star$ e-mail: marin@fi.infn.it}

\date{\today}

\begin{abstract}
A minimal observable length is a common feature of theories that aim to merge quantum physics and gravity. Quantum mechanically, this concept is associated to a nonzero minimal uncertainty in position measurements, which is encoded in deformed commutation relations. In spite of increasing theoretical interest, the subject suffers from the complete lack of dedicated experiments and bounds to the deformation parameters have just been extrapolated from indirect measurements. As recently proposed, low-energy mechanical oscillators could allow to reveal the effect of a modified commutator. Here we analyze the free evolution of high quality factor micro- and nano-oscillators, spanning a wide range of masses around the Planck mass $m_{\mathrm{P}}$ (${\approx 22\,\mu\mathrm{g}}$). The direct check against a model of deformed dynamics substantially lowers the previous limits on the parameters quantifying the commutator deformation.
\end{abstract}

\maketitle

The emergence of a minimal observable length, at least as small as the Planck length $L_{\mathrm{P}} =\sqrt{\hbar G/c^3} = 1.6 \times 10^{-35}$ m, is a general feature of different quantum gravity models \cite{garay,Hoss2012}.
In the framework of quantum mechanics, the measurement accuracy is at the heart of the Heisenberg relations, that, however, do not imply an absolute minimum uncertainty in the position. An arbitrarily precise measurement of the position of a particle is indeed possible at the cost of our knowledge about its momentum. This consideration motivated the introduction of generalized uncertainty principles (GUPs) \cite {veneziano,gross,garay,Hoss2012,amelino-camelia,maggiore1,scardigli,jizba}, such as
\begin{equation}
\Delta q \Delta p \geq \frac{\hbar}{2} \left(1+\beta_0 \left(\frac{L_{\mathrm{P}} \Delta p}{\hbar}\right)^2\right)  \qquad .
\label{eq1int}
\end{equation}
Eq. \ref{eq1int} implies indeed a nonzero minimal uncertainty $\Delta q_{min} = \sqrt{\beta_0} L_{\mathrm{P}}$. The dimensionless parameter $\beta_0$ is usually assumed to be around unity, in which case the corrections are negligible unless energies (lengths) are close to the Planck energy (length). However, since there are no theories supporting this assumption, the deformation parameter has necessarily to be bound by the experiments. Any experimental upper limit for $\beta_0 > 1$ would constrain new physics below the length scale $\sqrt{\beta_0} L_{\mathrm{P}}$ \cite{das}.

A direct consequence of relation (\ref{eq1int}) is an increase of the ground state energy $E_{min}$ of an harmonic oscillator. Recently, an upper limit to $E_{min}$ has been placed by analysing the residual motion of the first longitudinal mode of the bar detector of gravitational waves AURIGA \cite{auriga,aurigaNJP}. Although the imposed bound, $\beta_0 < 10^{33}$, is extremely far from the Planck scale, it provides a first measurement just below the electroweak scale (corresponding to ${10^{17} L_{\mathrm{P}}}$).

To the GUP (\ref{eq1int}) it is possible to associate a modified canonical commutator \cite{kempf}:
\begin{equation}
[q,p] = i \hbar \left(1+\beta_0 \left(\frac{L_{\mathrm{P}} \, p}{\hbar}\right)^2\right) \ .
\label{eq2int}
\end{equation}
Its introduction represents a further conceptual step, as it defines the algebraic structure from which the GUP should follow, and it implies changes in the whole energy
spectrum of quantum systems, as well as in the time evolution of a given observable.

Because of its importance as a prototype system, several studies have been focused on harmonic oscillators. Modifications of stationary states are calculated in Refs. \cite{kempf,chang1,lewis}. Approaches to construct generalized coherent states are proposed in Refs. \cite{ching,pedram1}. The modified time evolution and expectation values of position and momentum operators are discussed in \cite{nozari1,nozari2,pedram2}, while in Ref. \cite{chen} Chen {\it et al.} calculate the temporal behaviour of the position and momentum uncertainties in a coherent state, finding a squeezing effect.

In spite of this huge theoretical interest, the subject suffers from the complete lack of dedicated experiments and so far limits to the deformation parameters have been extrapolated from indirect measurements \cite{das,ali2,quesne}.
It has recently been proposed that the effect of a modified commutator could be revealed studying the opto-mechanical interaction of macroscopic mechanical oscillators \cite{pikovski}.
Here we elaborate a different experimental protocol and describe a set of dedicated experiments with state-of-the-art micro- and nano-oscillators.
We show that, in the Heisenberg picture of quantum mechanics and assuming the validity of the commutator (\ref{eq2int}) for the coordinates of the center-of-mass, the time evolution of its position exhibits an additional third harmonic term and a dependence of the oscillation frequency on its amplitude. The strength of such effects depends on $\beta_0$. We then analyze the dynamics of different oscillators in order to place upper bounds to the parameters quantifying the deformation to the standard quantum-mechanical commutator. Such bounds span a wide range of test mass values, around the landmark given by the Planck mass.
Previous limits, derived indirectly from the analysis of some metrological experiments, are substantially lowered, by several orders of magnitude.

\section{Results}

\subsection{Theoretical model}

The basic idea of our analysis is assuming that the commutation relations between the operator $q$ describing a measured position in a macroscopic harmonic oscillator, and its conjugate momentum $p$, are modified with respect to their standard form. In other words, and more generally, we suppose that the deformed commutator should be applied to any couple of position/momentum conjugate observables  that are treated in a quantum way in experiments and standard theories. At the same time, we keep the validity of the Heisenberg equations for the temporal evolution of an operator $\hat{O}$, i.e.
$\mathrm{d} \hat{O}/\mathrm{d}t =  [\hat{O},H]/i\hbar$,
where $H$ is the Hamiltonian. For an oscillator with mass $m$ and resonance angular frequency $\omega_0$, we also assume that the Hamiltonian maintains its classical form ${H = m \omega_0^2 q^2/2+ p^2/2 m}$. Such hypotheses are also underlying the proposal of Ref. \cite{pikovski}.

We first define the usual dimensionless coordinates $Q$ and $P$, according to ${q = \sqrt{\hbar/(m\omega_0)}\,Q}$ and ${p = \sqrt{\hbar m \omega_0}\,P}$. The Hamiltonian is now written in the standard form
$\, H = \frac{\hbar \omega_0}{2}\bigl(Q^2+P^2\bigr)\, $
and the commutator of Eq. (\ref{eq2int}) becomes
\begin{equation}
[Q,P] = i \bigl(1+\beta P^2\bigr),
\label{eq3}
\end{equation}
where
${\beta = \beta_0\,\bigl(\hbar m \omega_0/m_{\mathrm{P}}^2 c^2\bigr)}$ is a further dimensionless parameter that we assume to be small (${\beta \ll 1}$). Such assumption will have to be consistent with the experimental results.

We now apply the transform
\begin{equation}
P = \biggl(1+\frac{1}{3}\,\beta\tilde{P}^2 \biggr)\tilde{P}
\label{eq4}
\end{equation}
discussed, e.g., in Ref. \cite{quesne}. As we will see later, to our purpose $\tilde{P}$ is just an auxiliary operator, we do not need to decide if either $P$ or $\tilde{P}$ corresponds to the classical momentum. $Q$ and $\tilde{P}$ obey the (non deformed) canonical commutation relation
${[Q,\tilde{P}] = i}$. At the first order in $\beta$, the Hamiltonian can now be written as
\begin{equation}
H = \frac{\hbar \omega_0}{2}\Bigl(Q^2+\tilde{P}^2\Bigr)+\frac{\hbar \omega_0}{3}\,\beta \tilde{P}^4 \, .
\label{eq5}
\end{equation}

The Heisenberg evolution equations for $Q$ and $\tilde{P}$, using the Hamiltonian (\ref{eq5}) read
\begin{subequations}\label{eq6-7}
 \begin{eqnarray}
  \dot{Q} &=& \omega_0 \tilde{P}\,\biggl(1+\frac{4}{3}\,\beta \tilde{P}^2\biggr) \, , \label{eq6} \\
  \dot{\tilde{P}} &=& -\omega_0 Q \, . \label{eq7}
 \end{eqnarray}
\end{subequations}

The coupled relations (\ref{eq6-7}) are formally equivalent to the equations describing the evolution of a free anharmonic oscillator with position $ -\tilde{P}$ ($Q$ is its conjugate momentum), in a potential ${V = \omega_0^2\,\Bigl(\tilde{P}^2/2 + \beta\tilde{P}^4/3\Bigr)}$ containing a fourth-order component.

The Poincar\'e's solution \cite{Poincare}, for initial conditions ${-\tilde{P}(0) = A}$ and ${\dot{\tilde{P}}(0) = 0}$, is
$
-\tilde{P}(t) = A \bigl( (1+\epsilon/32) \cos (\tilde{\omega} t) \, - \, (\epsilon/32) \cos(3 \, \tilde{\omega} t)\bigr)\, \,
$
where ${\epsilon = - 4 A^2 \beta/3}$ and ${\tilde{\omega} = \Bigl(1-\tfrac{3}{8}\epsilon\Bigr)\,\omega_0}$. The solution is valid at the first order in $\epsilon$, and implies two relevant effects with respect to the harmonic oscillator: the appearance of the third harmonic and, less obvious, a dependence of the oscillation frequency on the amplitude (more precisely, a quadratic dependence of the frequency shift on the oscillation amplitude).

Using again Eq. (\ref{eq7}) to find $Q(t)$, keeping the first order in
$\beta Q_0^2$
where $Q_0$ is the oscillation amplitude for $Q$, we obtain
\begin{equation}
Q = Q_0 \biggl[\sin (\tilde{\omega} t) \, + \, \frac{\beta}{8} Q_0^2 \sin(3\, \tilde{\omega} t)\biggr] \, ,
\label{eq10}
\end{equation}
where
\begin{equation}
\tilde{\omega}=\biggl(1+\frac{\beta}{2}Q_0^2\biggr)\omega_0 \, .
\label{eq11}
\end{equation}
The position $Q(t)$ is our meaningful (i.e., measured) variable, whatever is the physical meaning of $P$ and $\tilde{P}$.
P.~Pedram calculates in Ref.~\cite{pedram2} the evolution of an harmonic oscillator with an Hamiltonian deformed according to the GUP considered in this work, and finds a frequency modified as (in our notation) ${\tilde{\omega}=\omega_0 \sqrt{1+\beta Q_0^2}}$. Such expression is equivalent to Eq. (\ref{eq11}) in the limit of small $({\beta Q_0^2})$, satisfied in the present work.

We have performed the experiments with highly isolated oscillators, i.e., with a high mechanical quality factor ${\mathcal{Q}_\mathrm{m} \coloneqq\omega_0 \tau}$, where $\tau$ is a long but nonetheless finite relaxation time, responsible for an additional term $-2P/\tau$ in the right hand side of Eq.~(\ref{eq7}). Damping has a twofold effect: (i) an exponential decay of the oscillation amplitude; (ii) a nontrivial \mbox{time-dependence} of the phase. In the limit ${\mathcal{Q}_\mathrm{m} \gg 1}$, the dynamics is described by a modified version of Eq.~(\ref{eq10}) with the replacements ${\tilde{\omega} t \to \Phi(t)}$, implying ${\tilde{\omega}(t)= \mathrm{d} \Phi / \mathrm{d} t}$, and ${Q_0 \to Q_0\,\exp(-t/\tau)}$. More details on the inclusion of damping in the evolution equations are reported in the Supplementary Material.

\begin{widetext}

\subsection{Experimental apparatus}

We have examined three kinds of oscillators, with masses of respectively $\approx 10^{-4}\,\mathrm{kg}$, ${\approx 10^{-7}\,\mathrm{kg}}$, and ${\approx 10^{-11}\,\mathrm{kg}}$. The measurements are performed by exciting an oscillation mode and monitoring a possible dependence of the oscillation frequency and shape (i.e., harmonic contents) on its amplitude, during the free decay. In order to keep a more general analysis, we will consider both indicators independently.

\begin{figure}[t!]
\includegraphics[width=0.9\linewidth]{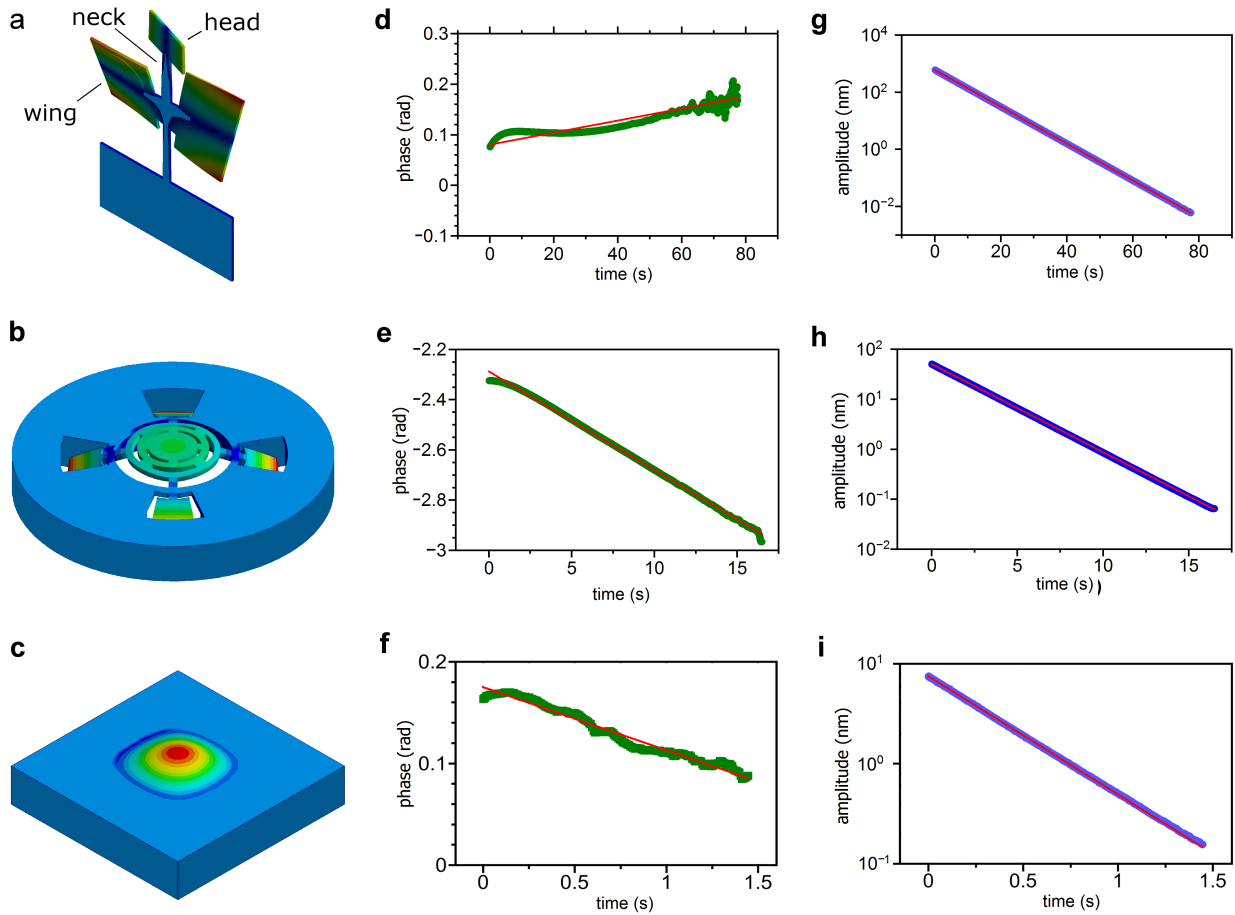}
\caption{Oscillating devices. Finite Elements simulation of the shapes of the oscillation modes investigated in this work (a, b, c), phase (d, e, f) and amplitude (g, h, i) of the oscillation during a free decay, obtained by phase-sensitive analysis of the measured position. In the left panels, the color scale represents the relative magnitude of the displacement for each modal shape, decreasing from red to blue. Red solid lines: linear and exponential fits respectively to the phase (blue dots) and the amplitude (green dots) experimental data. Graphs (a, d, g) refer to the DPO oscillator, consisting of two inertial members, head and a couple of wings, linked by a torsion rod (the neck) and  connected to the outer frame by a leg. The displayed AS mode consist of a twist of the neck around the symmetry axis and a synchronous oscillation of the wings.
The elastic energy is primarily located at the neck, where the maximum strain field occurs during the oscillations, while the leg remains at rest and the foot can be supported by the outer frame with negligible energy dissipation. Graphs (b, e, h) refer to the balanced wheel oscillator. The central disk has a diameter of 0.54 mm, and the shape of the beams maintain it flat during the motion (as shown by its homogeneous colour) reducing the dissipation on the 0.4 mm diameter optical coating. The four paddles are carefully sized in order to balance the stress induced by the strain of the beams on the supporting wheel, such that the joints correspond to nodal points. An additional external wheel further improves the isolation from the background. Graphs (c, g, i) refer to the $L = 0.5\,\mathrm{mm}$ side, $30\,\mathrm{nm}$ thick, square membrane of stoichiometric SiN membrane. Its high stress increases the mechanical quality factor thanks to the dilution effect.}
\end{figure}

The first device is a ``double paddle oscillator'' (DPO) \cite{spiel01} made from a ${300\,\mu\mathrm{m}}$ thick silicon plate (Fig. 1a).
Thanks to its shape, for two particular balanced oscillation modes, the Antisymmetric torsion modes (AS), the oscillator is supported by the outer frame with negligible energy dissipation and it can therefore be considered as isolated from the background \cite{borrielli11}.
Vibrations are excited and detected capacitively, thanks to two gold electrodes evaporated over the oscillator, and two external electrodes.
The sample is kept in a vacuum chamber, and its temperature is stabilized at ${293\,\mathrm{K}}$ within ${2\,\mathrm{mK}}$, a crucial  feature to maintain a constant resonance frequency during the measurements.
We have monitored the AS2 mode, with a resonance frequency of ${5636\,\mathrm{Hz}}$ and a mechanical quality factor of ${1.18 \times 10^5}$ (at room temperature).
The overall center-of-mass (c.m.) of the oscillator remains at rest during the AS motion. To our purpose, we consider the positions of the couple of c.m.'s corresponding to the two half-oscillators that move symmetrically around the oscillator rest plane (a deeper discussion of this issue is reported in Ref.~\cite{auriga}). The meaningful mass is the reduced mass of the couple of half-oscillators, that is calculated by FEM simulations and is ${m = 0.033\,\mathrm{g}}$.

For the measurements at intermediate mass we have used a silicon wheel oscillator, made on the ${70\,\mu\mathrm{m}}$ thick device layer of a SOI wafer and composed of a central disk kept by structured beams \cite{SerraAPL2012}, balanced by four counterweights on the beams joints that so become nodal points (Fig. 1b) \cite{Borrielli2014}. On the surface of the central disk, a multilayer \mbox{${\mathrm{SiO}_2}$/${\mathrm{Ta}_2\mathrm{O}_5}$} dielectric coating forms an high reflectivity mirror.  The device also includes intermediate stages of mechanical  isolation. The design strategy
allows to obtain a balanced oscillating mode (its resonance frequency is ${141\,797\,\mathrm{Hz}}$), with a planar motion of the central mass (significantly reducing the contribution of the optical coating to the structural dissipation) and a strong isolation from the frame.
The oscillator is mechanically excited using a piezoelectric ceramic glued on the sample mount. The surface of the core of the device works as end mirror in one arm of a stabilized Michelson interferometer, that allows to measure its displacement.
The quality factor surpasses $10^6$ at the temperature of ${4.3\,\mathrm{K}}$, kept during the measurements.
As for the DPO, the c.m. of the oscillator remains at rest and, for the following analysis of the possible quantum gravity effects, we consider the reduced mass ${m=20\,\mu\mathrm{g}}$. We have also performed room temperature measurements on a simpler device, lacking of counterweights, with an oscillating mass of ${77\,\mu\mathrm{g}}$ and a frequency of ${128\,965\,\mathrm{Hz}}$.

Finally, the lighter oscillators is a ${L = 0.5\,\mathrm{mm}}$ side, ${30\,\mathrm{nm}}$ thick, square membrane of stoichiometric silicon nitride, grown on a ${5\,\mathrm{mm} \times 5\,\mathrm{mm}}$, ${200\,\mu\mathrm{m}}$ thick silicon substrate~\cite{Harris}. Thanks to the high tensile stress, the vibration can be described by standard membrane modes, the lowest one (monitored in this work) with shape ${z(x,y) = A \cos(\pi x/L) \cos (\pi y/L)}$ where ${(x,y)}$ are the coordinates measured from the membrane center, along directions parallel to its sides (Fig. 1c). The physical mass of the membrane is ${20\,\mathrm{ng}}$, respectively, and the c.m. is at the position ${(0,0,z_{\mathrm{cm}})}$ with ${z_{\mathrm{cm}} = 4 A/\pi^2}$ (the central position $A$ is the monitored observable).
We have performed the measurements in a cryostat at the temperature of ${65\,\mathrm{K}}$ and  pressure of ${10^{-4}\,\mathrm{Pa}}$, where the oscillation frequency is ${747\,\mathrm{kHz}}$
and the quality factor is ${8.6\times 10^5}$. Excitation and readout are performed as in the experiment with the wheel oscillators.

\subsection{Measurements and data analysis}

The first step in the data analysis is applying to the data stream $q(t)$ a numerical lock-in: the two quadratures $X(t)$ and $Y(t)$ are calculated by multiplying the data respectively by ${\sin(\omega_0 t)}$ and ${\cos (\omega_0 t)}$, where $\omega_0$ is the oscillation angular frequency of the acquired time series, estimated preliminarily from a spectrum, and applying appropriate low-pass filtering. The oscillation amplitude is calculated as ${q_0(t) = \sqrt{X^2 + Y^2}}$ and the phase as ${\Phi(t) = \arctan(Y/X)}$. For the DPO oscillator, this process is directly performed by the hardware lock-in amplifier, that is
also used to frequency down-shift the signal of the wheel oscillator at cryogenic temperature before its acquisition. ${q_0(t)}$ is fitted with an exponential decay (examples are shown in Fig.~1) while ${\Phi(t)}$, that always remains within ${\pm \pi\,\mathrm{rad}}$, is fitted with a linear function that gives the optimal frequency and phase with respect to the preliminary tries $\omega_0$ and ${\Phi(0)=0}$. The residuals
${\Delta\Phi}$ of the fit are differentiated to estimate the fluctuations $\Delta \omega$ in the oscillation frequency.
In Fig. 2 we show $\Delta\omega$ as a function of $q_0$, together with its fit  with the function ${\Delta\omega = a + b q_0^2}$. The derived value and uncertainty in the quadratic coefficient $b$ are the meaningful quantities that can be used to establish upper limits to the deformation parameter $\beta_0$. The background mechanical noise in all the experiments is dominated by the oscillator thermal noise (as verified with spectra taken without excitation). The consequent statistical uncertainty in the calculated $\Delta\omega$ is inversely proportional to the amplitude $q_0$, and such a weight is indeed used in the fitting procedures.

\begin{figure}[t!]
\includegraphics[width=0.9\linewidth]{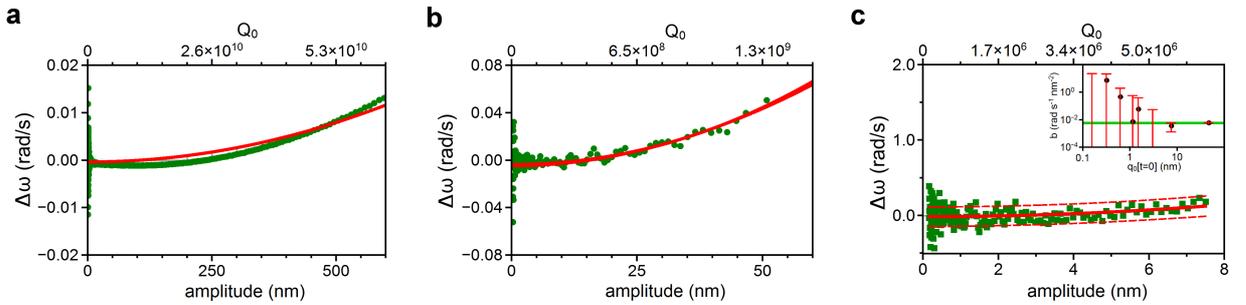}
\caption{Residual angular frequency fluctuations as a function of the oscillation amplitude. The fluctuations $\Delta \omega$ are measured during the free decay, for the DPO (a), wheel (b) and membrane (c) oscillators. On the upper axes, the same oscillation amplitudes are normalized to the respective oscillator ground state wavefunction width $\sqrt{\hbar/m \omega_0}$. Red solid lines are the fits with Equation \ref{eq11}, dashed lines reports the 95$\%$ confidence area. In the inset, we report the values of the quadratic coefficient $b$ measured for the membrane oscillator at different excitation amplitudes, with their 95$\%$ confidence error bars (for appreciating the improvement in the accuracy, we just show the positive vertical semi-axis in logarithmic scale). For the two points at highest amplitude, the measured $b$ is significantly different from zero. The green lines show the interval of $b$ calculated from the nonlinear behaviour observed in the frequency domain for stronger excitation.}
\end{figure}

In the case of the DPO oscillator (Fig. 2a), $\Delta\omega$ vs $q_0$ has a clear shape that is given by the intrinsic oscillator non-linearity. A similar, weaker effect is observed for the wheel oscillator at cryogenic temperature and for the membrane, at the largest excitation amplitudes (Fig. 2b-c). The quadratic coefficient for the membrane is in agreement with its calculation based on the nonlinear behaviour observed for larger amplitudes in the frequency domain \cite{Harris2}. Since structural nonlinearity is hard to model, we cannot distinguish it from possible quantum gravity effects. Therefore, we just place our upper bound in correspondence of the first (the strongest) nonlinear behaviour encountered.
The meaningful quantity to calculate an upper limit to $\beta_0$ is the mean value of $b$ plus its uncertainty. The latter is calculated from the standard deviation on several independent measurements, and it is in agreement with the error estimated from the residuals of each fit, after decimation of the data sets to obtain uncorrelated data points. The experiment has been repeated for several excitation levels, finding the expected improvements in the upper limit to $b$ at increasing amplitudes (inset of Fig. 2c).

As previously discussed, a further useful indicator is the amplitude of the third harmonic component, also extracted with a lock-in procedure. The value of $\beta$ inferred from such parameter using Eq.~(\ref{eq10}), is to ascribe to the relatively poor linearity of the readout, and is therefore considered as an upper limit to possible quantum gravity effects.

A model-independent constraint to possible effects of a deformed commutator can be derived from the residual frequency fluctuations $\Delta\omega$, considered as a function of the oscillation amplitude (reported in Fig. 2). To this purpose, we summarize in Table I the maximum relative frequency shift and the maximum dimensionless oscillation amplitude ${Q_0(0)}$, for the different oscillator masses examined in this work. These data can be used to test any modified dynamics and provide the consequent upper limits to the involved parameters.

\begin{table}
\begin{tabular}{cccccccc}
\toprule
Mass  & Frequency  & Max.~ampl. & Max.~$Q_0$ & Max.~${\Delta\omega/\omega_0}$ & ${\beta}$ & ${\beta_0}$ & indicator\\
(${\mathrm{kg}}$) & (${\mathrm{Hz}}$) & (${\mathrm{nm}}$) & & & & &\\
\hline
${3.3 \times 10^{-5}}$ & ${5.64 \times 10^3}$ & ${600}$ & ${6\times 10^{10}}$ &  ${4 \times 10^{-7}}$ & ${7 \times 10^{-29}}$ & ${3 \times 10^7}$ & ${\Delta\omega}$\\
\textquotedbl & \textquotedbl &&&& ${7 \times 10^{-25}}$ & ${2 \times 10^{11}}$ &  ${3^{\mathrm{rd}}}$~harmonic\\
${7.7 \times 10^{-8}}$ & ${1.29 \times 10^5}$ &&&& ${8 \times 10^{-24}}$ & ${5 \times 10^{13}}$ &  ${\Delta\omega}$ \\
\textquotedbl & \textquotedbl & &&&${2 \times 10^{-19}}$ & ${2 \times 10^{18}}$ & ${3^{\mathrm{rd}}}$~harmonic \\
${2 \times 10^{-8}}$ & ${1.42 \times 10^5}$ & ${55}$ & ${7\times 10^{8}}$ &  ${6 \times 10^{-8}}$ & ${3 \times 10^{-25}}$ & ${6 \times 10^{12}}$ & ${\Delta\omega}$ \\
${2 \times 10^{-11}}$ & ${7.47 \times 10^5}$ & ${7.5}$ & ${7\times 10^{6}}$ &  ${4 \times 10^{-8}}$& ${4 \times 10^{-21}}$ & ${2 \times 10^{19}}$ & ${\Delta\omega}$\\
\textquotedbl & \textquotedbl & ${47}$ & ${4\times 10^{7}}$ &  ${3 \times 10^{-6}}$& \textquotedbl & \textquotedbl & ${\Delta\omega}$\\
\textquotedbl & \textquotedbl & &&&${2 \times 10^{-14}}$ & ${1 \times 10^{26}}$ &  ${3^{\mathrm{rd}}}$~harmonic \\
\end{tabular}
\caption{Results of the experiment. Maximum relative frequency shifts measured for different oscillators, corresponding oscillation amplitudes, and upper limits to the deformation parameters $\beta$ and $\beta_0$ obtained in this work.}
\end{table}

\end{widetext}

For a more accurate and specific bound, we focus on the model described by Eqs.~(\ref{eq10}-\ref{eq11}). The values and uncertainties in $\beta$ and $\beta_0$ are obtained from $b$ and from the third harmonic distortion, using the oscillator parameters (namely, its mass and frequency). In Table I we summarize our results for the different upper limits, given at the $95\%$ confidence level. The results for $\beta_0$ are also displayed in Fig. 3 as a function of the oscillator mass, and compared with some previously existing limits. We have achieved a significant improvement, by many orders of magnitude, working on  systems with disparate mass scales and considering different measured observables.

\begin{figure}[t!]
\includegraphics[width=0.9\linewidth]{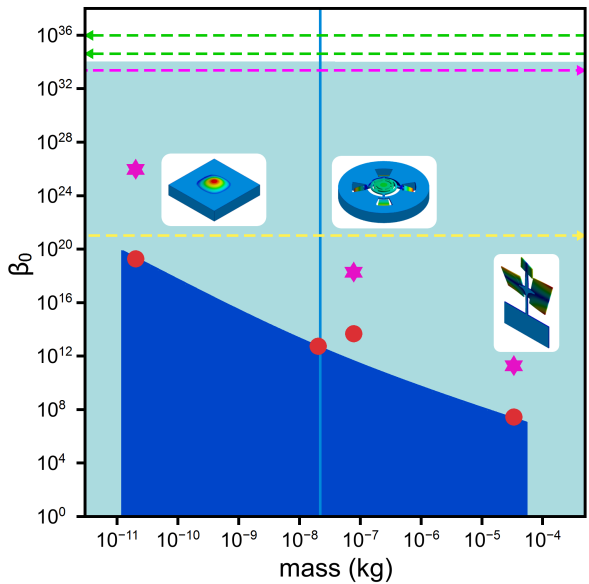}
\caption{Upper limits to the deformed commutator. The parameter $\beta_0$ quantifies the deformation to the standard commutator between position and momentum, or the scale $\sqrt{\beta_0}L_{\mathrm{P}}$ below which new physics could come into play. Full symbols reports its upper limits obtained in this work, as a function of the mass. Red dots: from the dependence of the oscillation frequency from its amplitude; magenta stars: from the third harmonic distortion. In the former data set, for the intermediate mass range (10 $\div$ 100 $\mu$g), we report the results obtained with two different oscillators. Light blue shows the area below the electroweak scale, dark blue the area that remains unexplored. Dashed lines reports some previously estimated upper limits, obtained in mass ranges outside this graph (as indicated by the arrows). Green: from high resolution spectroscopy on the hydrogen atom, considering the ground state Lamb shift (upper line) \cite{ali2} and the 1S-2S level difference (lower line) \cite{quesne}. Magenta: from the AURIGA detector \cite{auriga,aurigaNJP}. Yellow: from the lack of violation of the equivalence principle \cite{ghosh}. The vertical line corresponds to the Planck mass (22 $\mu$g).}
\end{figure}

\section{Discussion}

We have performed an extended experimental analysis of the possible dependence of the oscillation frequency and third harmonic distortion on the oscillation amplitude in micro- and nano-oscillators, spanning a wide range of masses. Assuming that a deformed commutator between position and momentum governs the dynamics through standard Heisenberg equations, we obtain a reduction by many orders of magnitude of the previous upper limits to the parameters quantifying the commutator deformation.
We remark that the measurements have been performed on state of the art oscillators, allowing low statistical uncertainty (due to the high mechanical quality factor), low background noise (thanks to the shot-noise limited detection and the cryogenic environment), high frequency stability (beyond the resonance linewidth), and the highest excitation amplitude allowed by each oscillator. The latter condition is not commonly explored in metrological micro- and nano-oscillators \cite{Nguyen2007,Galliou2013}, and we could indeed achieve the limit given by the intrinsic oscillators non-linearity. These effects are not well mastered at present, therefore we have kept the conservative attitude of setting an upper limit to the overall nonlinear behavior, which includes possible quantum gravity effects. A detailed modeling of the structural nonlinearity could allow in the future to subtract their effects from our data (in particular in the case of the DPO, for which the shape of $\Delta \omega$ vs $q_0$ is clearly different from a parabola), and thus set even stronger limits to the remaining nonlinearity and actually to $\beta_0$.
The mentioned crucial properties (high $\mathcal{Q}_\mathrm{m}$, high resonance frequency at a given mass, high frequency stability) must be conserved or improved in possible further experiments aiming to lower the bounds on the deformation parameters. We stress however that, at present, the wall in our experiment is given by dynamic range, actually determined by the structural nonlinearity that should be reduced to improve the results. In this regard, an interesting possibility to be explored is the use of high quality bulk crystalline resonators  \cite{Galliou2013,Bourhill2015}.

Extending the use of the Heisenberg evolution equations with deformed commutators from an ideal particle to a macroscopic dynamics is not free from conceptual problems~\cite{Hossenfelder}.

A direct extrapolation from quantum to classical dynamics, discussed e.g. in Refs. \cite{chang2,nozari3,pedram3}, implies crucial consequences, the first being the violation of the equivalence principle \cite{Ali2011,ghosh,scardigli2}. Current bounds to such violation, obtained using sensitive torsion balances \cite{adelberger}, correspond to a deformation parameter $\beta_0 < 10^{21}$ \cite{ghosh}.
Our limits are substantially lower. Our model shows that remarkable deviations from classical trajectories are, in any case, expected as soon as the momentum is of the order of (or exceeds) $m_{\mathrm{P}} c$. This condition is straightforward at astronomic level \cite{Maziashvili2012}, and even for kg scale bodies. This may either indicate the breakdown of the Eherenfest' theorem at all scales \cite{nozari2} (requiring to revise the rules connecting quantum to classical dynamics), or a possible mass dependence of the deformation parameter. In this context our experiments, involving a wide range of masses, taking as ``natural'' reference the Planck mass, become particularly meaningful. Our experimental results, when used to set limits on the deformed commutator (\ref{eq2int}), should not be simply intended as a check of possible deformations of quantum mechanics, but as a test of a 'composite' hypothesis, involving also the form of the classical limit corresponding to the modified quantum rules.

As a second general remark, it should be underlined that the role of the centre-of-mass coordinates in a deformed space is still a matter of debate. As recently remarked in Ref. \cite{Amelino-Camelia2}, the motion of the centre-of-mass typically do not involve a Planck energy concentrated in a Planck-scale volume.
In the same article, the author constructs a deformed commutator for a composite system starting from a number $N$ of elementary constituents, and shows that the deformation parameter should scale as $N^{-2}$. Therefore, even assuming the constituent particles to be affected by Planck scale physics, the centre-of-mass of a composite macroscopic body would be much more weakly affected. This approach leads to the interesting (maybe troubling) conclusion that free elementary particles should feel quantum gravitational effects in a different way with respect to e.g. protons or atoms or, in other words, that spacetime properties should depend on the kind of particle~\cite{quesne,Amelino-Camelia}.
We further remark that, at the present stage, we don't know at which constituent-particle-level quantum gravity effects could intervene~\cite{Amelino-Camelia}, and there are no theories even suggesting what such "elementary constituents" should be. Other works suggest instead that the effects should scale as the number of elementary interactions \cite{Amelino2011}.

A different point of view is to consider the effects of an intrinsically discrete spacetime on the dynamics of a quantum system. We remark that, in quantum mechanics, the wavefunction associated to the centre-of-mass has properties that cannot be simply reduced to the coordinates of the constituent particles. It has been shown that a discretization of spacetime, (e.g. related to the creation and annihilation of particle-antiparticle pairs) would naturally suggest discretization of the Hilbert space associated to the considered quantum system \cite{Buniy}. Although our universe might still be infinite in extent, any experiment or observation involves just a finite region of spacetime. In Ref. \cite{Marchiolli} is investigated the emergence of extended uncertainty relations for discrete coordinate and momentum operators in such finite discrete configuration spaces, which can be formulated in the form of a GUP. In this context, deformed commutators appear as the manifestation of the background discreteness, with the minimal scale being a fundamental property of spacetime. In this case, one could expect also the low-energy motion of a macroscopic body to be affected, independently on the measurement process.

While all these approaches are formally correct, they rely on very different hypothesis, whose validity should be checked by experimental measurements. This represents a strong motivation for the realization of experiments involving macroscopic mechanical oscillators. We notice that clear quantum signatures have been recently obtained even in ``macroscopic'' nano-oscillators \cite{oconnell10,teufel11,chan11,safavi12} very similar to those exploited in this work, suggesting that well isolated mechanical oscillators are indeed privileged experimental systems to explore the classical-to-quantum transition.
Since gravity effects could play a role in the wavefunction decoherence that marks such transition \cite{Diosi,Penrose}, and it cannot be excluded that quantum gravity is inextricably linked to peculiar quantum features, an intriguing extension of the present experiment (or a similar investigation) would naturally be performed with macroscopic oscillators in a fully quantum regime.

\begin{widetext}

\section{Supplementary material: Adding damping to the evolution in the presence of deformed commutators}

\subsection{Equations for a damped oscillator}

Our analysis is based on two basic simple and reasonable assumptions: i) time evolution of a system is generated by its Hamiltonian, and therefore it is dictated by the Heisenberg equation; ii) the Hamiltonian of an harmonic oscillator with position $q$ and conjugated momentum $p$ is the usual one, ${H = m \omega_0^2 q^2/2+ p^2/2 m}$. The first assumption is motivated by the connection between time translations and a system Hamiltonian, while the second assumption is motivated by an overwhelming set of experimental results. Therefore the only effect of the existence of a minimal length is associated with a deformation of the commutator between position and momentum of the harmonic oscillator,
\begin{equation}
[q,p] = i \hbar \biggl[1+\beta_0 \biggl(\frac{p}{m_{\mathrm{P}} c}\biggr)^2\biggr],
\label{eq1}
\end{equation}
where $m_{\mathrm{P}}$ is the Planck mass (${\approx 22\,\mu\mathrm{g}}$), and $\beta_0$ is a dimensionless parameter that we are limiting with the experiment.

We first define the usual dimensionless coordinates $Q$ and $P$, according to ${q = \sqrt{\hbar/(m\omega_0)}\,Q}$ and ${p = \sqrt{\hbar m \omega_0}\,P}$. The Hamiltonian is now written in the standard form
\begin{equation}
H_{\cal S} = \frac{\hbar \omega_0}{2}\bigl(Q^2+P^2\bigr)
\label{eq2}
\end{equation}
and the commutator in Eq. (\ref{eq1}) becomes
\begin{equation}
[Q,P] = i \bigl(1+\beta P^2\bigr),
\label{eq3}
\end{equation}
where
${\beta = \beta_0\,\bigl(\hbar m \omega_0/m_{\mathrm{P}}^2 c^2\bigr)}$ is a further dimensionless parameter that we assume to be small (${\beta \ll 1}$). Such assumption will have to be consistent with the experimental results.

Damping and noise acting on the resonator are due to the coupling to an external environment which, as extensively discussed in the literature~\cite{caldleg2,qnoise,ocon,GIOV01}, can be modeled in terms of a set of independent harmonic oscillators, with frequency $\omega_j$, couplings
$k_j$, and whose canonical coordinates $q_j$ and $p_j$ have been again rescaled as we have done with the canonical variables of the resonator of interest. The total system Hamiltonian is
\cite{caldleg2,qnoise,ocon}
\begin{equation}
H_{\cal S} + \sum_{j}\frac{\hbar \omega_j}{2}\left[p_{j}^2+\left(q_j-k_j Q\right)^2\right],
\label{primo1}
\end{equation}
and the quantum Langevin equations for the operators $Q$ and $P$ can be obtained starting from the Heisenberg
equations for $q(t)$, $p(t)$ associated with this Hamiltonian. We assume the usual commutation rules for the reservoir oscillators $[q_j,p_k]=i\delta_{jk}$, i.e., that they are \emph{not} modified by minimum length arguments, and we will justify this choice later on.

The corresponding evolution equations are
\begin{eqnarray}
  \dot{Q}(t) &=& \frac{i}{\hbar} \left[H_{\cal S},Q(t)\right]\label{secondo1} \\
  \dot{P}(t) &=& \frac{i}{\hbar} \left[H_{\cal S},P(t)\right]
  +\sum_{j}\omega_j k_{j}\left[q_{j}(t)-k_{j}Q(t)\right]
  \label{secondo2}\\
  \ddot{q_{j}}(t) &=& - \omega_{j}^2 \left[q_{j}(t) - k_{j}Q(t)\right] \; . \label{secondo3}
\end{eqnarray}
We now proceed in the usual way by formally solving the dynamics of the reservoir oscillators $q_j(t)$,
\begin{equation}\label{formsol}
    q_j(t)=q_j(t)^{(0)}+k_j\int_0^t ds \omega_j \sin\omega_j(t-s) Q(s),
\end{equation}
where $q_j(t)^{(0)}=q_j(0)\cos\omega_j t +p_j(0)\sin\omega_j t$ is the free dynamics of each oscillator. We now integrate by parts the integral on the right hand side and rewrite Eq.~(\ref{formsol}) as
\begin{equation}\label{formsol2}
    q_j(t)-k_j Q(t)=q_j(t)^{(0)}-k_j \cos\omega_j t Q(0) -\int_0^t ds k_j \cos\omega_j(t-s) \dot{Q}(s).
\end{equation}
We now insert this formal solution into Eq.~(\ref{secondo2}), so that we finally get the general Langevin equations for the resonator of interest,
\begin{eqnarray}
  \dot{Q}(t) &=& \frac{i}{\hbar} \left[H_{\cal S},Q(t)\right]\label{secondo1bis} \\
  \dot{P}(t) &=& \frac{i}{\hbar} \left[H_{\cal S},P(t)\right] - K(t)Q(0)-\int_0^t ds K(t-s) \dot{Q}(s) + F(t),
\end{eqnarray}
where $K(t)=\sum_j \omega_j k_j^2 \cos\omega_j t$ is the reservoir memory kernel function and $F(t)=\sum_j k_j \omega_j q_j(t)^{(0)}$ is a reservoir operator which must be interpreted as a thermal stochastic force operator with zero mean value.

As it is well known, irreversible damping due to the reservoir is
obtained only when an infinite number of oscillators, distributed
over a continuum of frequencies, is considered. The simplest option is to perform the continuous limit
according to the usual Markovian prescription of a flat spectral distribution,
\begin{equation}
\sum_{j} \omega_j k_{j}^{2} \cdots \rightarrow \int_{0}^{+\infty} d \omega \,
\omega k^{2}(\omega) \, \frac{ d n }{ d \omega} \cdots \,= \,\frac{\gamma}{\omega_0 \pi}\int_{0}^{+\infty} d \omega \cdots,
\label{primo15}
\end{equation}
where $dn/d\omega $ is the oscillators density, and $\gamma $ is
just the damping rate of the resonator, which is associated with the overall coupling strength between the oscillator and its reservoir. With this choice, we have $K(t)=(\gamma/\omega_0)\delta(t)$, where $\delta(t)$ is the Dirac delta function, so that the Heisenberg equations for the oscillator of interest
become
\begin{eqnarray}
  \dot{Q}(t) &=& \frac{i}{\hbar} \left[H_{\cal S},Q(t)\right] \label{sblib}\\
  \dot{P}(t) &=& \frac{i}{\hbar} \left[H_{\cal S},P(t)\right] -\frac{\gamma}{\omega_0} \dot{Q}(t) + F(t) \; .
\label{secondo5}
\end{eqnarray}
We can now justify why we have assumed unmodified commutators for the reservoir oscillators. These oscillators are fictitious systems describing unspecified excitations of the reservoir, affecting the dynamics of our system of interest only through the spectral density of the coupling coefficients, determining the damping force acting on it. Therefore, it is not evident at all that modified commutators must be considered for such fictitious objects, and even when assuming modified commutators, their overall effect on the reservoir spectral density and on the damping rate $\gamma$ would be typically negligible.

We now introduce the auxiliary operator $\tilde{P}$ related to the momentum $P$ by the equation
\begin{equation}
P = \biggl(1+\frac{1}{3}\,\beta\tilde{P}^2 \biggr)\tilde{P}
\label{eq4}
\end{equation}
which, as shown in Ref. \cite{quesne}, is such that $Q$ and $\tilde{P}$ obey the usual (non deformed) canonical commutation relation
${[Q,\tilde{P}] = i}$ at first order in $\beta$. Therefore, at first order in $\beta$, the equation of motion for $Q$ and $\tilde{P}$ become
\begin{subequations}\label{eq6-7}
 \begin{eqnarray}
  \dot{Q} &=& \omega_0 \tilde{P}\,\biggl(1+\frac{4}{3}\,\beta \tilde{P}^2\biggr), \label{eq6} \\
  \dot{\tilde{P}} &=& -\omega_0 Q -\frac{\gamma}{\omega_0} \dot{Q} \biggl(1-\beta \tilde{P}^2\biggr), \label{eq7}
 \end{eqnarray}
\end{subequations}
where we have also neglected the thermal noise term because we will consider the evolution of mean values from now on. By taking the derivative of Eq.~(\ref{eq7}) and using Eq.~(\ref{eq6}), we get the following equation for the auxiliary variable $\tilde{P}$
\begin{equation}\label{eqptilde}
    \ddot{\tilde{P}}+\gamma \dot{\tilde{P}}\left(1+\beta \tilde{P}^2\right)+\omega_0^2 \tilde{P}+\frac{4}{3}\beta\omega_0^2 \tilde{P}^3=0.
\end{equation}
The physical quantity of interest, which is the one really measured in our experiments, is however $Q(t)$, which is obtained  from the knowledge of $\tilde{P}(t)$ stemming from the solution of Eq.~(\ref{eqptilde}), by formally solving Eq.~(\ref{eq6}) for $Q(t)$ which, assuming the initial condition $Q(0)=0$, gives
\begin{equation}\label{eqq}
    Q(t)=\omega_0\int_0^t ds \tilde{P}(s)\biggl(1+\frac{4}{3}\,\beta \tilde{P}^2(s)\biggr).
\end{equation}

\subsection{Approximate solution of the evolution equation}

We have now to solve Eq.~(\ref{eqptilde}) at first order in $\beta$ and also exploiting the weak damping condition  ${\cal Q}_m \gg 1$. We adopt a multiple scale approach in which the small nonlinear terms at first order in $\beta$ introduce two first order modifications to the damped oscillatory solution at zeroth order in $\beta$ of Eq.~(\ref{eqptilde}): i) a slowly varying amplitude of the zeroth order solution; ii) addition of a small third harmonic component.
The trial solution is therefore
\begin{equation}\label{eqtrial}
    \tilde{P}(t)= A_1(t)e^{\lambda t}+A_3(t)e^{3\lambda t}+ {\rm c.c},
\end{equation}
where $\lambda =-\gamma/2+i\omega_1$, with $\omega_1=\sqrt{\omega_0^2-\gamma^2/4}$, is the complex eigenvalue associated with the zeroth-order linear equation, and $A_3(t)$ is at first order in $\beta$. Separating the terms with damped oscillations at the frequency $\omega_1$ from those oscillating at $3 \omega_1$, we get the following two equations for $A_1(t)$ and $A_3(t)$ (again at first order in $\beta$):
\begin{eqnarray}\label{a1}
  && \ddot{A}_1(t)+i 2 \omega_1 \dot{A}_1(t)+4\beta \omega_0^2 A_1(t)|A_1(t)|^2 e^{-\gamma t} + \gamma \beta\left[2|A_1(t)|^2\left(\dot{A}_1(t)+\lambda A_1(t)\right)+A_1(t)^2\left(\dot{A}_1^*(t)+\lambda^* A_1^*(t)\right)\right]e^{-\gamma t}=0 , \\
  && \ddot{A}_3(t)+(6 \lambda+\gamma) \dot{A}_3(t)+\left(9\lambda^2+3\lambda \gamma+\omega_0^2\right)A_3(t)+\frac{4}{3}\beta \omega_0^2 A_1^3(t)  + \gamma \beta\left(\dot{A}_1(t)+\lambda A_1(t)\right)A_1(t)^2=0.\label{a2}
\end{eqnarray}
One has to solve Eq.~(\ref{a1}) for $A_1(t)$ and then insert this solution into Eq.~(\ref{a2}) in order to get the solution for $A_3(t)$. Since $A_1(t)$ is slowly varying we can neglect the second order derivative $\ddot{A}_1(t)$ in Eq.~(\ref{a1}); moreover for typical values of the mechanical quality factor, the last term on the left hand side of Eq.~(\ref{a1}), is much smaller that the others. Under these two assumptions, which are well justified in the parameter regime of our experiments, Eq.~(\ref{a1}) can be easily solved because $|A_1(t)|^2$ is a constant of motion, and one gets
\begin{equation}\label{a1sol}
    A_1(t)=A_1(0)\exp\left[2i \frac{\beta \omega_0^2}{\omega_1 \gamma}\left(1-e^{-\gamma t}\right)|A_1(0)|^2\right].
\end{equation}
For what concerns the solution for $A_3(t)$ we notice that Eq.~(\ref{a2}) is an inhomogeneous second order linear differential equation for $A_3(t)$ with constant coefficients, in which the last two terms in the left hand side depending upon $A_1(t)$ act as driving terms. Therefore the solution is the sum of two components: i) the general solution of the associated homogeneous equation;  ii) a particular solution of the inhomogeneous equation. However the first component, when multiplied by $e^{3\lambda t}$ will only provide a small correction to the term proportional to $e^{\lambda t}$, i.e., a negligible correction to $A_1(t)$, and therefore the relevant solution is only the second component. At first order in $\beta$ such particular solution is the slowly varying solution obtained by neglecting both $\ddot{A}_3(t)$ and $\dot{A}_3(t)$. Neglecting again the last term on the left hand side of Eq.~(\ref{a2}) proportional to $\gamma \beta$ due to large mechanical quality factor, we finally get
\begin{equation}\label{a2sol}
    A_3(t)=-\frac{2\beta \omega_0^2}{3\lambda (4 \lambda +\gamma)} A_1^3(t) \;\;\simeq \frac{\beta }{6} A_1^3(t)\;\;{\rm in\;the\;limit\;of\;high\;} {\cal Q}_m.
\end{equation}
In Eq.~(\ref{a2sol}) we have used the fact that $\omega_0^2=-\lambda^2-\gamma \lambda $ in the denominator. If we now insert Eqs.~(\ref{a1sol})-(\ref{a2sol}) into the trial solution of Eq.~(\ref{eqtrial}) and the resulting expression into Eq.~(\ref{eqq}), at the first order in $\beta$ the position $Q(t)$ can be written as
\begin{equation}\label{eqq1}
    Q(t)=\omega_0\int_0^t ds \, A_1(s)e^{\lambda s}+\left( A_3(s)+\frac{4}{3}\,\beta A_1^3(s) \right) e^{3\lambda s}+ {\rm c.c} \;\;\simeq \omega_0\int_0^t ds \, A_1(s)e^{\lambda s}+\frac{3}{2}\,\beta A_1^3(s) \, e^{3\lambda s}+ {\rm c.c.} \, .
\end{equation}
By taking $A_1(s)$ and $A_1^3(s)$ out of the integral as slowing varying envelops and performing the integration on $e^{\lambda s}$ and $e^{3\lambda s}$, defining $Q_0 = 2 A_1(0)$ and $\tau = 2/\gamma$ (amplitude decay time), we obtain
\begin{equation}\label{eqq2}
    Q(t)=\frac{\omega_0}{2 \lambda} Q_0 \,e^{-\frac{t}{\tau}}
\left[ e^{i \Phi(t)} + \frac{\beta}{8} Q_0^2 \,e^{-2\frac{t}{\tau}}e^{i 3 \Phi(t)} \right]+ {\rm c.c.}
\end{equation}
where the time-dependent phase $\Phi(t)$ is
\begin{equation}\label{eqphi}
\Phi(t) = \frac{\beta}{2} \frac{\omega_0^2}{\omega_1} \frac{1-e^{-\gamma t}}{\gamma} Q_0^2 + \omega_1 t = \int_0^t ds \,
\omega_1 \left[ 1+\frac{\beta}{2}\frac{\omega_0^2}{\omega_1^2} \, Q_0^2 \, e^{-2\frac{s}{\tau}} \right]
\end{equation}
and, again in the limit of high ${\cal Q}_m$ (i.e., $\omega_0/\omega_1 \simeq 1$ and $\omega_0/\lambda \simeq 1/i$), we find the expression described in the main text.

We have performed a numerical integration of Eqs.~(\ref{eq6-7}) with our typical experimental parameters, and we have indeed found that our final expression of $Q(t)$, used for the analysis of the experimental data, well reproduces the simulation.

\end{widetext}


\begin{thebibliography}{99}
\bibitem{garay} Garay, L. G. Quantum gravity and minimum length. {\it Int. J. Mod. Phys. A} {\bf 10}, 145-165 (1995).
\bibitem{Hoss2012} Hossenfelder, S. Minimal length scale scenarios for quantum gravity. {\it Living Rev. Relativity} {\bf 16}, 2 (2013).
\bibitem{veneziano} Amati, D., Ciafaloni, M. \& Veneziano, G. Superstring collisions at planckian energies. {\it Phys. Lett. B} {\bf 197}, 81-88 (1987).
\bibitem{gross} Gross, D. J. \& Mende, P. F. String theory beyond the Planck scale. {\it Nucl. Phys. B} {\bf 303}, 407-454 (1988).
\bibitem{maggiore1} Maggiore, M. A generalized uncertainty principle in quantum gravity.
{\it Phys. Lett. B} {\bf 304}, 65-69 (1993).
\bibitem{scardigli} Scardigli, F. Generalized uncertainty principle in quantum gravity from micro-black hole gedanken experiment. {\it Phys. Lett. B} {\bf 452}, 39-44 (1999).
\bibitem{amelino-camelia} Amelino-Camelia, G. Doubly special relativity: First results and key open problems. {\it Int. J. Mod. Phys. D} {\bf 11}, 1643-1669 (2002).
\bibitem{jizba} Jizba, P., Kleinert, H. \& Scardigli, F. Uncertainty relation on a world crystal and its applications to micro black holes. {\it Phys. Rev. D} {\bf 81}, 084030 (2010).
\bibitem{das} Das, S. \& Vagenas, E. C. Universality of quantum gravity corrections. {\it Phys. Rev. Lett. } {\bf 101}, 221301 (2008).
\bibitem{auriga} Marin, F. {\it et al.} Gravitational bar detectors set limits to Planck-scale physics on macroscopic variables. {\it Nature Phys.} {\bf 9}, 71-73 (2013).
\bibitem{aurigaNJP} Marin, F. {\it et al.} Investigation on Planck scale physics by the AURIGA
gravitational bar detector. {\it New J. Phys.} {\bf 16}, 085012 (2014).
\bibitem{kempf} Kempf, A. Uncertainty relation in quantum mechanics with quantum group symmetry. {\it J. Math. Phys.} {\bf 35}, 4483-4496 (1994); Kempf, A., Mangano, G. $\&$ Mann, R. B. Hilbert space representation of the minimal length uncertainty relation. {\it Phys. Rev. D} {\bf 52}, 1108-1118 (1995).
\bibitem{chang1} Chang, L.N., Minic, D., Okamura, N. \& Takeuchi, T. Exact solution of the harmonic oscillator in arbitrary dimensions with minimal length uncertainty relations. {\it Phys. Rev. D} {\bf 65} 125027 (2002).
\bibitem{lewis} Lewis, Z. \&  Takeuchi, T. Position and momentum uncertainties of the normal and inverted harmonic oscillators under the minimal length uncertainty relation. {\it Phys. Rev. D} {\bf 84}, 105029 (2011).
\bibitem{ching} Ching, C.L. \& Ng, W.K. Generalized coherent states under deformed quantum mechanics with maximum momentum. {\it Phys. Rev. D} {\bf 88} 084009 (2013).
\bibitem{pedram1} Pedram, P. Coherent States in Gravitational Quantum Mechanics. {\it Int. J. Mod. Phys. D} {\bf 22} 1350004 (2013).
\bibitem{nozari1} Nozari, K. Some aspects of Planck scale quantum optics. {\it Phys. Lett. B} {\bf 629}, 41-52 (2005).
\bibitem{nozari2} Nozari, K. \& Azizi, T. Gravitational induced uncertainty and dynamics of harmonic oscillator {\it Gen. Relativ. Gravit.} {\bf 38} 325-331 (2006).
\bibitem{pedram2} Pedram, P. New approach to nonperturbative quantum mechanics with minimal length uncertainty. {\it Phys. Rev. D} {\bf 85}, 024016 (2012).
\bibitem{chen} Chen, Y.-Y., Feng, X.-L., Oh, C.H. \& Xu, Z.-Z. Squeezing effect induced by minimal length uncertainty. Preprint at http://arxiv.org/abs/1405.4655 (2014).
\bibitem{ali2} Ali, A. F., Das, S. \& Vagenas, E. C. A proposal for testing quantum gravity in the lab. {\it Phys. Rev. D} {\bf 84}, 044013 (2011).
\bibitem{quesne} Quesne, C. \& Tkachuk, V. M. Composite system in deformed space with minimal length. {\it Phys. Rev. A} {\bf 81}, 012106 (2010).
\bibitem{pikovski} Pikovski, I., Vanner M. R., Aspelmeyer, M., Kim, M. S. \& Brukner, $\check{\mathrm{C}}$. Probing Planck-scale physics with quantum optics, {\it Nature Phys. } {\bf 8}, 393-397 (2012).
\bibitem{Poincare} Strogatz, S. H., Nonlinear Dynamics and Chaos: With Applications to Physics, Biology, Chemistry, and Engineering (Westview, Cambridge MA, 1994).
\bibitem{spiel01} Spiel, C. L., Pohl, R. O. \& Zehnder, A. T.
Normal modes of a Si(100) double-paddle oscillator.
{\it Rev. Sci. Inst.} {\bf 72}, 1482-1491 (2001).
\bibitem{borrielli11} Borrielli, A., Bonaldi, M., Serra, E., Bagolini, A., \& Conti, L.
Wideband mechanical response of a high-Q silicon double-paddle oscillator.
{\it J. Micromech. Microeng.} {\bf 21}, 065019 (2011).
\bibitem{SerraAPL2012} Serra, E. {\it et al.}
A  "low-deformation mirror" micro-oscillator with ultra-low optical and mechanical losses.
{\it Appl. Phys. Lett.} {\bf 101}, 071101 (2012).
\bibitem{Borrielli2014} Borrielli, A. {\it et al.}
Design of silicon micro-resonators with low mechanical and optical losses for quantum optics experiments.
{\it Microsyst. Technol.} {\bf 20}, 907-917 (2014).
\bibitem{Harris} Thompson, J. D., Zwickl, B. M., Jayich, A. M., Marquardt, F., Girvin, S. M.  \&  Harris, J. G. E.
Strong dispersive coupling of a high-finesse cavity to a micromechanical membrane.
{\it Nature} {\bf 452}, 72-75 (2008).
\bibitem{Harris2} Zwickl, B. M., Shanks, W. E., Jayich, A. M., Yang, C., Bleszynski Jayich, A. C., Thompson, J. D. \& Harris, J. G. E.
High quality mechanical and optical properties of commercial silicon
nitride membranes.
{\it Appl. Phys. Lett.} {\bf 92}, 103125 (2008).
\bibitem{Nguyen2007} Nguyen, C. T.-C. MEMS Technology for timing and frequency control. {\it IEEE Transactions on Ultrasonics, ferroelectrics, and frequency control} {\bf 54}, 251-270 (2007).
\bibitem{Galliou2013} Galliou, S., Goryachec, M., Bourquin, R., Abb\'e, P., Aubry, J. P. \& Tobar, M.E. {\it Scientific Reports} {\bf 3}, 2132 (2013).
\bibitem{Bourhill2015} Bourhill, J., Ivanov, E.,
\& Tobar, M.E. Precision Measurement of a low-loss Cylindrical Dumbbell-Shaped Sapphire
Mechanical Oscillator using Radiation Pressure. Preprint at http://arxiv.org/abs/1502.07155 (2015).
\bibitem{Hossenfelder} Hossenfelder, S. {\it SIGMA} {\bf 10}, 074 (2014).
\bibitem{chang2} Benczik, S. {\it et al.} Short distance versus long distance physics: The classical limit of the minimal length uncertainty relation. {\it Phys. Rev. D} {\bf 66}, 026003 (2002).
\bibitem{nozari3} Nozari, K. \&  Akhshabi, S. On the stability of planetary circular orbits in
noncommutative spaces. {\it Chaos Solitons Fractals} {\bf 37}, 324-331 (2008).
\bibitem{pedram3} Pedram, P. A higher order GUP with minimal length uncertainty and maximal momentum II: Applications. {\it Phys. Lett. B} {\bf 718}, 638-645 (2012).
\bibitem{Ali2011} Ali, A. F. Minimal length in quantum gravity, equivalence principle and holographic entropy bound. {\it Class. Quantum Grav.} {\bf 28}, 065013 (2011).
\bibitem{ghosh} Ghosh, S. Quantum gravity effects in geodesic motion and predictions of equivalence principle violation. {\it Class. Quantum Grav.} {\bf 31}, 025025 (2014).
\bibitem{scardigli2} Scardigli, F. \& Casadio, R. Gravitational tests of the Generalized Uncertainty Principle. Preprint at http://arxiv.org/abs/1407.0113 (2014).
\bibitem{adelberger} Schlamminger, S., Choi, K.Y., Wagner, T.A., Gundlach, J.H., \& Adelberger, E.G. Test of the equivalence principle using a rotating torsion balance. {\it Phys. Rev. Lett.} {\bf 100}, 041101 (2008).
\bibitem{Maziashvili2012} Maziashvili, M. \& Megrelidze, L. Minimum-length deformed quantum mechanics/quantum field theory, issues, and problems. {\it Prog. Theor. Exp. Phys.} {\bf 2013}, 123B06 (2013).
\bibitem{Amelino-Camelia2} Amelino-Camelia, G. Challenge to Macroscopic Probes of Quantum Spacetime Based on Noncommutative Geometry. {\it Phys. Rev. Lett.} {\bf 111}, 101301 (2013).
\bibitem{Amelino-Camelia} Amelino-Camelia, G. Planck-scale soccer-ball problem: a case of mistaken identity. Preprint at http://arxiv.org/abs/1407.7891 (2014).
\bibitem{Amelino2011} Amelino-Camelia,G., Freidel, L., Kowalski-Glikman, J. \& Smolin, L., Relative locality and the soccer ball problem, {\it Phys. Rev. D} {\bf 84}, 087702 (2011).
\bibitem{Buniy} Buniy, R. V., Hsu, S. D. H. \& Zee, A. Is Hilbert space discrete? {Phys. Lett. B} {\bf 630}, 68-72 (2005).
\bibitem{Marchiolli} Marchiolli, M. A. \& Ruzzi, M. Theoretical formulation of finite-dimensional discrete phase spaces: I. Algebraic structures and uncertainty principles. {Ann. Phys.} {\bf 327}, 1538-1561 (2012).
\bibitem{oconnell10} O'Connell, A. D. {\it et al.} Quantum ground state and single-phonon control of a mechanical resonator. {\it Nature} {\bf 464}, 697-713 (2010).
\bibitem{teufel11} Teufel, J. D. {\it et al.}
Sideband cooling of micromechanical motion to the quantum ground state. {\it Nature} {\bf 475}, 359-363 (2011).
\bibitem{chan11} Chan, J.{\it et al.}
Laser cooling of a nanomechanical oscillator into its quantum ground state. {\it Nature} {\bf 478}, 89-92 (2011).
\bibitem{safavi12} Safavi-Naeini, A. H. {\it et al.} Observation of quantum motion of a nanomechanical resonator. {\it Phys. Rev. Lett. } {\bf 108}, 033602 (2012).
\bibitem{Diosi} Di\'osi, L. A universal master equation for the gravitational violation of quantum mechanics. {\it Phys. Lett. A} {\bf 120}, 377-381 (1987).
\bibitem{Penrose} Penrose, R. On gravity's role in quantum state reduction. {\it Gen. Rel. Grav.} {\bf 28}, 581-600 (1996).

\bibitem{caldleg2}A.O. Caldeira and A.J. Leggett, Ann. Phys. (N.Y.) {\bf
149}, 374 (1983).
\bibitem{qnoise}C. W. Gardiner and P. Zoller, {\em Quantum Noise} (Springer-Verlag,
Berlin, 2004), Chap. 3.
\bibitem{ocon}G.W. Ford, J.T. Lewis, and R.F. O'Connell, Phys. Rev. A
{\bf 37}, 4419 (1988).
\bibitem{GIOV01}
V. Giovannetti, D. Vitali, Phys. Rev. A, \textbf{63}, 023812 (2001).

\end{thebibliography}
\end{document}